\documentclass[journal, hidelinks]{IEEEtran}

\pdfoutput=1
\IEEEoverridecommandlockouts

\usepackage{acronym}
\usepackage{amsmath}
\usepackage{amssymb}
\usepackage[english]{babel}
\usepackage{cite}
\usepackage[shortlabels,inline]{enumitem}
\usepackage{etoolbox}
\usepackage{graphicx}
\usepackage{hyperref}
\usepackage{bookmark}
\usepackage{ifthen}
\usepackage{cleveref}
\usepackage[utf8]{inputenc}
\usepackage{lipsum}
\usepackage{siunitx}
\usepackage[dvipsnames]{xcolor}
\usepackage[T1]{fontenc}
\usepackage[subnum]{cases}
\usepackage{tikz-cd}
\usepackage{epstopdf}
\graphicspath{{./fig/pbs_results/}}
\usepackage{bm}
\usepackage{array}
\usepackage{subfig}
\usepackage{filecontents}

\allowdisplaybreaks

\newcommand{\Figure}[1]{Fig.~\ref{#1}}
\newcommand{\Figures}[2]{Figs.~\ref{#1} and~\ref{#2}}
\newcommand{\FigureList}[2]{Figs.~\ref{#1} --~\ref{#2}}

\newcommand{\Section}[1]{Section~\ref{#1}}

\newcommand{\Sections}[2]{Sections~\ref{#1}~and~\ref{#2}}

\acrodef{MC}{molecular communication}
\acrodef{OOK}{ON-OFF keying}
\acrodef{ISI}{inter-symbol interference}
\acrodef{BER}{bit error rate}
\acrodef{TX}{transmitter}
\acrodef{TXs}{transmitters}
\acrodef{RX}{receiver}
\acrodef{RXs}{receivers}
\acrodef{w.r.t.}{with respect to}
\acrodef{GFP}{green fluorescent protein}
\acrodef{EX}{eraser}
\acrodef{SM}{signaling molecule}
\acrodef{SMs}{signaling molecules}

\makeatletter
\long\def\@makecaption#1#2{\ifx\@captype\@IEEEtablestring%
    \footnotesize\begin{center}{\normalfont\footnotesize #1}\\
        {\normalfont\footnotesize\scshape #2}\end{center}%
    \@IEEEtablecaptionsepspace
    \else
    \@IEEEfigurecaptionsepspace
    \setbox\@tempboxa\hbox{\normalfont\footnotesize {#1.}~~ #2}%
    \ifdim \wd\@tempboxa >\hsize%
    \setbox\@tempboxa\hbox{\normalfont\footnotesize {#1.}~~ }%
    \parbox[t]{\hsize}{\normalfont\footnotesize \noindent\unhbox\@tempboxa#2}%
    \else
    \hbox to\hsize{\normalfont\footnotesize\hfil\box\@tempboxa\hfil}\fi\fi}
\makeatother

\newcommand{\scaleSection}{\vspace{-0.2cm}}
\newcommand{\scaleSubsection}{\vspace{-0.1cm}}

\newcommand{\scaleSectionBelow}{\vspace{-0.1cm}}
\newcommand{\scaleSubsectionBelow}{\vspace{-0.1cm}}

\makeatother

\begin{document}
\title{Switchable Signaling Molecules for Media Modulation: Fundamentals, Applications, and Research Directions}
\author{\IEEEauthorblockN{Lukas Brand, Maike Scherer, Sebastian Lotter, Teena tom Dieck, \\Maximilian Schäfer, Andreas Burkovski, Heinrich Sticht, Kathrin Castiglione, and Robert Schober}\vspace{-0.3cm}}
\maketitle

\begin{abstract}
Although visionary applications of molecular communication (MC), such as long-term continuous health monitoring by cooperative in-body nanomachines, have been proposed, MC is still in its infancy when it comes to practical implementation. In particular, long-term experiments and applications face issues such as depletion of signaling molecules (SMs) at the transmitter (TX) and inter-symbol interference (ISI) at the receiver (RX). To overcome these practical challenges, a new class of SMs with switchable states seems to be promising for future MC applications.
In this work, we provide an overview of existing switchable SMs, and classify them according to their properties. Furthermore, we highlight how switchable SMs can be utilized as information carriers for media modulation. In addition, we present theoretical and experimental results for an end-to-end MC system employing the \textit{green fluorescent protein variant "Dreiklang"} (GFPD) as switchable SM. Our experimental results show, for the first time, successful information transmission in a closed-loop pipe system using media modulation. Finally, we discuss media modulation specific challenges and opportunities.
\end{abstract}
\setlength{\belowdisplayskip}{2pt}
\setlength{\belowdisplayshortskip}{2pt}
\acresetall
\vspace{0.2cm}

\scaleSection
\section{Introduction}\label{intro}
\scaleSectionBelow
Communication systems employing optical and electrical signals for information transmission have evolved tremendously over the past decades, creating today's global communication network. However, these well-established communication techniques face challenges when it comes to realizing synthetic micro- and nano-scale communication, e.g., within the human body. On the contrary, \textit{natural} communication mechanisms based on molecular signals exist from macro- to nano-scale, e.g., hormone signaling between organs via the blood stream and synaptic signaling in the nervous system, which has motivated the emerging field of \textit{synthetic} \ac{MC}.

One of the main differences between \ac{MC} and conventional, electromagnetic forms of communication is the adoption of \ac{SMs}\acused{SM} as information carriers. Despite the key role that \ac{SMs} play in \ac{MC}, \ac{SMs} are usually modeled as generic molecules in theoretical synthetic \ac{MC} studies \cite{jamali2019channel}.
In contrast, in natural \ac{MC} systems, a large variety of highly specialized \ac{SMs} is employed. For example ions, such as $\mathrm{Na}^+$, $\mathrm{K}^+$, $\mathrm{Cl}^-$, and $\mathrm{Ca}^{2+}$, are specialized \ac{SMs} tailored for signal transmission and reception controlled by ion channels \cite{frank2005overview}.

Additionally, despite their large variety, natural \ac{MC} systems always operate in a resource efficient manner, which they often achieve by reutilization of \ac{SMs}. Well-known natural examples for this resource efficiency include the reversible switching in epigenetics \cite{holliday2006epigenetics}, protein phosphorylation \cite{jin2012modular}, and redox reactions \cite{schieber2014ros}, which are also employed in the human body.
In epigenetics, a DNA sequence can be interpreted as a natural \ac{SM}. Epigenetic changes enable the switching of gene activity between \textit{on} and \textit{off} by methylation without altering the DNA sequence, i.e., the expression of \textit{different} phenotypes at \textit{different} instants of time can be communicated by using and re-using the \textit{same} DNA sequence over and over again. Similarly, via protein phosphorylation, enzymes, which are natural \ac{SMs} acting as biological catalysts, are activated or deactivated reversibly rather than disposed after the signaling. In redox reactions, an acceptor or donor either "steals" an electron from or donates an electron to available \ac{SMs} in the surroundings and thereby generates oxidizing agents or reducers, i.e., only the electric charge of the \ac{SMs} is altered, which is resource and energy efficient.

These examples reveal that nature uses a broad toolkit to adapt the properties, which in the following we refer to as states, of \textit{permanently available} molecules.
If not needed, these molecules are switched off until they are required for a certain task, i.e., when not needed they are inactive, but not degraded.

\begin{figure*}[!tbp]
  \centering
  \begin{minipage}[t]{1\linewidth}
    \centering
    \resizebox{135mm}{!}{
    \fontsize{5pt}{5pt}\selectfont
    \def\svgwidth{3.333in}
    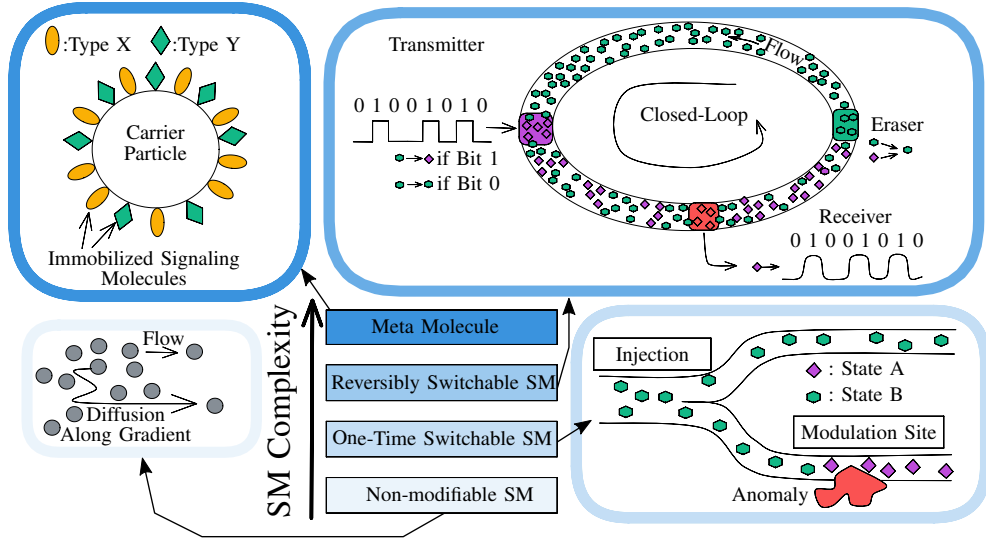}
    \caption{Classification of \ac{SMs}, which vary in terms of their functional versatility. More sophisticated \ac{SMs} may enable groundbreaking applications such as continuous health monitoring (lower right image) and long-term information transmission by \ac{MC} (upper right image), but are also more complex to design and synthesize.}\label{fig:level_development}
  \end{minipage}
  \vspace{-0.35cm}
\end{figure*}

Inspired by these natural processes, the use of \ac{SMs} with adjustable properties is promising for synthetic \ac{MC} systems, as we illustrate in the following.
In conventional synthetic \ac{MC} systems, the \ac{SMs} are stored, released, and then have to be replenished by the \ac{TX}. In practice, the replenishment of the \ac{SMs} at the \ac{TX} is difficult if not impossible to achieve, especially in medical and, in general, microscale applications.
Hence, it is desirable to develop systems employing novel \ac{TX} concepts and modulation techniques to overcome this practical challenge.
We argue that switchable \ac{SMs} are well suited for this purpose and pave the way to facilitate long-term \ac{MC} applications such as continuous health monitoring.

The use of switchable \ac{SMs} allows for writing information into the state of \ac{SMs} already present in the medium. Thus, this process is referred to as \textit{media modulation}, as introduced in \cite{Brand2022MediaModulation}.
Media modulation belongs to the family of index modulation schemes \cite{Basar2019}.
Examples of index modulation in conventional electromagnetic wave-based communication are the activation of one out of several \ac{TX} antennas and altering the wireless channel fading characteristics, which are commonly referred to as spatial modulation and media modulation \cite{Basar2019}, respectively. In the latter case, after modulating the carrier signal, additional information is embedded into the transmit signal by adjustable mirrors at the \ac{TX}.
As the information is embedded into the \textit{transmitted signal, which may already carry some information and exists in the medium}, the term media modulation has been coined. Similarly, for \ac{MC} systems, the process of modulating the states of \textit{switchable molecules, which already exist in the medium}, is also denoted as media modulation \cite{Brand2022MediaModulation}.
Hereby, the switching from one state to another is triggered by an external electromagnetic (including optical) or chemical stimulus. Depending on the type, switchable \ac{SMs} enable data storage, i.e., they can be written once or are reversibly writable, which makes them interesting for information transmission tasks.

This magazine article generalizes the media modulation concept introduced first in \cite{Brand2022MediaModulation} as follows. First, we discuss switchable \ac{SMs} as a promising general class of \ac{SMs} for \ac{MC} and review existing \ac{SMs}. Second, we provide a conceptional classification of \ac{SMs}, which is based on the functional versatility of the \ac{SMs} in \Section{ReviewAndFramework}.
Third, in \Section{MediaModulationSection}, we first highlight the benefits of media modulation based on switchable \ac{SMs} by comparing it to a conventional \ac{MC} system releasing molecules from a point \ac{TX} via particle-based simulations (PBSs).
Fourth, we present in detail two selected switchable \ac{SMs}, namely \textit{flip green fluorescent protein} (FlipGFP), which is introduced to the \ac{MC} community for the first time, and the \textit{green fluorescent protein variant "Dreiklang"} (GFPD), which we previously introduced in \cite{Brand2022MediaModulation}. For these two \ac{SMs}, we highlight possible applications, for which the in-depth understanding of the media modulation processes, i.e., the switching of the states of the \ac{SMs} for conveying information, is needed.
Fifth, we discuss theoretical results for the communication performance of an \ac{MC} system employing GFPD as switchable \ac{SM} in \Section{subsec:TheoreticalResults}.
To demonstrate that media modulation based \ac{MC} is practically feasible, we provide first experimental results for an end-to-end \ac{MC} link in \Section{subsec:practicalResults}.
Finally, we discuss design challenges that media modulation based \ac{MC} faces in \Section{sec:Challenges}, which motivate new research directions, before concluding with \Section{sec:conclusion}.

\scaleSection
\section{Framework for Classifying Signaling Molecules}\label{ReviewAndFramework}
\scaleSectionBelow
Several different models for \ac{TX}, channel, and \ac{RX} have been proposed for various applications of \ac{MC}.
However, usually the \ac{SMs} considered in \ac{MC} systems are generic, whereas for certain applications, more elaborate \ac{SMs} are desired.
For example, for long-term health monitoring, \ac{SMs} could be designed such that disease markers switch the states of the \ac{SMs}, i.e., the health status is reflected in the state of the \ac{SMs}.
In the following, we introduce a conceptual classification of \ac{SMs} based on their functional versatility, which also reflects their complexity, cf. \Figure{fig:level_development}. We argue that the most basic function of an \ac{SM} is being a particle with non-changeable properties once released into the channel. As a more complex type of \ac{SMs}, we identify switchable molecules, which are either one-time or reversibly switchable, and are discussed in \Sections{blankStateSignalingMolecules}{ReWritableSignalingMolecules}, respectively. Finally, in \Section{MetaSignalingMolecules}, we propose functionalized carrier particles, which we denote as meta molecules, as the most sophisticated type of \ac{SMs}. Meta molecules combine different switchable molecules and their properties in one entity.

\scaleSubsection
\subsection{One-Time Switchable SMs}\label{blankStateSignalingMolecules}
\scaleSubsectionBelow
The final state of one-time switchable molecules is not yet fixed when they are released into the channel. In particular, these \ac{SMs} are injected in a blank state, e.g., state $X$, propagate in the channel, and may be switched by an appropriate stimulus at the modulation site. Depending on the application, such stimuli are actively controlled, e.g., light stimuli for synthetic information transmission, or exist passively, e.g., locally concentrated markers of a disease, which stimulate nearby \ac{SMs}, cf. lower right part of \Figure{fig:level_development}. When triggered, the \ac{SMs} switch from the default state $X$ to another state, e.g., state $Y$, i.e., $X \, \rightarrow Y$. After being modulated by the \ac{TX}, the state is final and cannot be changed again. The switching changes the properties of the \ac{SMs}, which can be exploited by the \ac{RX} to read out the state.
In summary, switchable \ac{SMs} enable spatial and temporal decoupling of the molecule injection and the information modulation process, which is essential for in-body applications where the location of modulation is hard to access or unknown. FlipGFP is an example of a one-time switchable \ac{SM} and will be discussed in detail in \Section{FlipGFP}.

\scaleSubsection
\subsection{Reversibly Switchable SMs}\label{ReWritableSignalingMolecules}
\scaleSubsectionBelow
While the one-time switchable feature enables "write-once, read many times" memory behavior, some \ac{SMs} allow for a reset of the switching, i.e., they can be re-written multiple times, which we refer to as reversibly switchable. In particular, reversibly switchable \ac{SMs} allow to write, read, and delete information repeatedly. This evens the path for closed-loop MC applications, where \ac{SMs} can be reused many times. In fact, we envision reversibly switchable \ac{SMs} to be the key to realize closed-loop \ac{MC} systems, where replenishment of \ac{SMs} is difficult and \ac{ISI} mitigation is critical. In closed-loop MC applications, \ac{TX} and \ac{RX} can be complemented by an \ac{EX} which returns the \ac{SMs} to the ground state. Thus, the information from past transmissions is deleted to allow for another modulation by the \ac{TX}, cf. upper right part of \Figure{fig:level_development}.

Being able to erase information may be also beneficial in terms of security, where certain information should not leave a given region, i.e., should be indecipherable outside this region. This might be also helpful, when a larger system is partitioned into subsystems, similar to the cellular system in classic mobile communication, and cross-talk is to be minimized.

GFPD is an example of a reversibly switchable \ac{SM} and will be discussed in detail in \Section{GFPD_intro}. The theoretical and practical feasibility of \ac{MC} using GFPD is validated in \Section{GFPD_theory_and_practice}.
\scaleSubsection
\subsection{Meta Molecules: Molecule Type Integration on Carrier Particles}\label{MetaSignalingMolecules}
\scaleSubsectionBelow

For some applications, it is desirable to have several functionalities merged in one \ac{SM}, which we refer to as a meta molecule. These molecules integrate different switchable molecules and act as carrier, cf. upper left part of \Figure{fig:level_development}.
Due to their composition, meta molecules can be switched between more than two different states. Hence, utilizing such meta molecules may enable the implementation of higher-order molecule shift keying modulation schemes, where the information is encoded by the type of integrated molecules, which is switched, cf. Type X and Type Y in the upper left part of \Figure{fig:level_development}.

One example for synthetic meta molecules are polymersomes, i.e., engineered vesicles, which can exhibit multiple different properties by embedding proteins, e.g., \acp{GFP}, into their membrane \cite{klermund2016simple}. The immobilization process may stabilize the proteins. Thus, polymersomes may facilitate longer lifetimes of proteins which is desired for long-term applications. Furthermore, varying the size of polymersomes allows altering their propagation behavior, as size impacts random movement due to Brownian motion.

In summary, meta molecules enable higher-order modulation schemes and prolong the life-time of the attached proteins.
\scaleSection
\section{Media Modulation based MC}\label{MediaModulationSection}
\scaleSectionBelow
In this section, we explain the advantages of media modulation based MC with a toy example, discuss two existing switchable \ac{SMs} for media modulation, and present results obtained from theoretical work and an experimental testbed.

\subsection{Advantages of Media Modulation Based MC in a Nutshell}\label{CompareBothSystems}
\begin{figure*}[!tbp]
    \begin{minipage}[t]{1\textwidth}
      \centering
      \resizebox{180mm}{!}{
      \fontsize{8pt}{11pt}\selectfont
      \def\svgwidth{500px}
      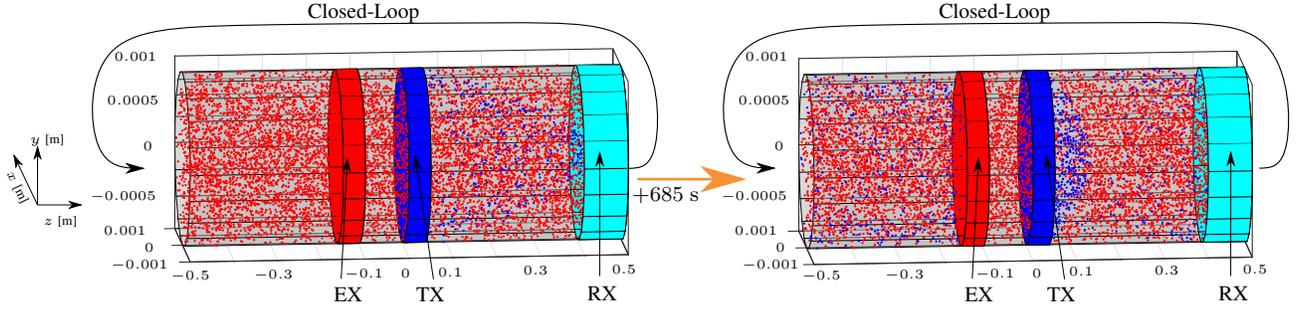}\vspace{-0.3cm}
      \caption{PBS snapshots at simulation times $t= 20 \, \si{\second}$ (left) and $t= 705 \, \si{\second}$ (right) of an \ac{MC} system using media modulation based \ac{OOK} for information transmission.}\label{fig:PBS_MM}
    \end{minipage}
    \vspace{-0.3cm}
  \end{figure*}

\begin{figure*}[!tbp]
      \begin{minipage}[t]{1\textwidth}
        \centering
        \resizebox{180mm}{!}{
        \fontsize{8pt}{11pt}\selectfont
        \def\svgwidth{500px}
        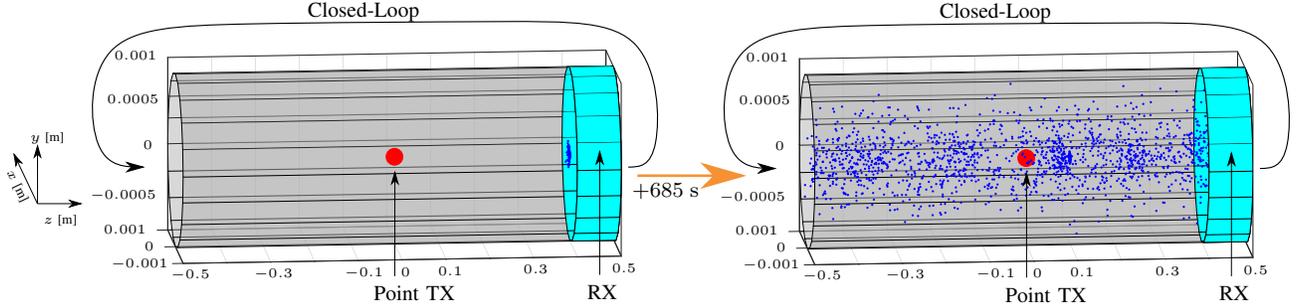}\vspace{-0.3cm}
        \caption{PBS snapshots at simulation times $t= 20 \, \si{\second}$ (left) and $t= 705 \, \si{\second}$ (right) of a conventional \ac{MC} system, which uses \ac{OOK} by emitting molecules from a point \ac{TX} for information transmission.}\label{fig:PBS_classic}
      \end{minipage}
      \vspace{-0.3cm}
    \end{figure*}

In \FigureList{fig:PBS_MM}{subfig-5}, we compare a media modulation system, which uses switchable \ac{SMs} already present in the channel, to a conventional \ac{MC} system with a molecule emitting point \ac{TX}, which uses basic, non-switchable \ac{SMs}. For this comparison, we conduct PBSs, simulating a closed-loop cylindrical duct filled with a fluid.
Snapshots of the PBSs are shown in \Figures{fig:PBS_MM}{fig:PBS_classic}.
The parameter values are chosen such that they are in the same order of magnitude as those of the human cardiovascular system \cite[Chap. 14]{hall2020guyton}. The fluid exhibits laminar flow along the $z$-axis with a maximum velocity of $v=0.02 \,\si{\meter \per \second}$ at the center of the duct. Thus, the \ac{SMs} propagate due to both diffusion, characterized by diffusion coefficient $D= 1 \,\times \,10^{-10} \, \si{\meter \squared \per \second}$, and background flow.
We assume a symbol duration of $T = 20 \, \si{\second}$.
The \acp{TX} in both systems use \ac{OOK}, a form of binary unipolar Amplitude Shift Keying (ASK), i.e., the \acp{TX} are active for bit $1$ and inactive for bit $0$ transmissions, respectively. In particular, in the beginning of a bit $1$ transmission interval, the \ac{TX} triggers the switching of the molecules from state B (red) to state A (blue) within its volume (blue cylinder) in the media modulation system, cf. \Figure{fig:PBS_MM}, and releases $N = 100$ \ac{SMs} in the conventional system from a dedicated point (large red dot), cf. \Figure{fig:PBS_classic}, respectively.
For the media modulation system, we assume that $N_{\mathrm{Sys}}=10000$ switchable \ac{SMs} are initially uniformly distributed in the system.
These \ac{SMs} are initially in state B.
The \ac{EX} (red cylinder), cf. \Figure{fig:PBS_MM}, is always active and switches molecules from state A to state B, when they pass through.
The \ac{RX} is modeled as transparent. We assume that the \acp{RX} in both systems are able to perfectly count the blue \ac{SMs} within their volumes (cyan cylinders), while the red \ac{SMs} are not visible to the \acp{RX}. Hence, the information is conveyed by the blue \ac{SMs}. Nevertheless, for media modulation, the red \ac{SMs} are shown for completeness, as the \ac{TX} relies on the existence of enough \ac{SMs} in state B to switch them to state A for bit 1 transmissions.
In \Figure{fig:PBS_MM}, the laminar flow profile is apparent from the distribution of the blue \ac{SMs}.

\begin{figure}[!tbp]
    \centering
        \includegraphics[width=1\columnwidth]{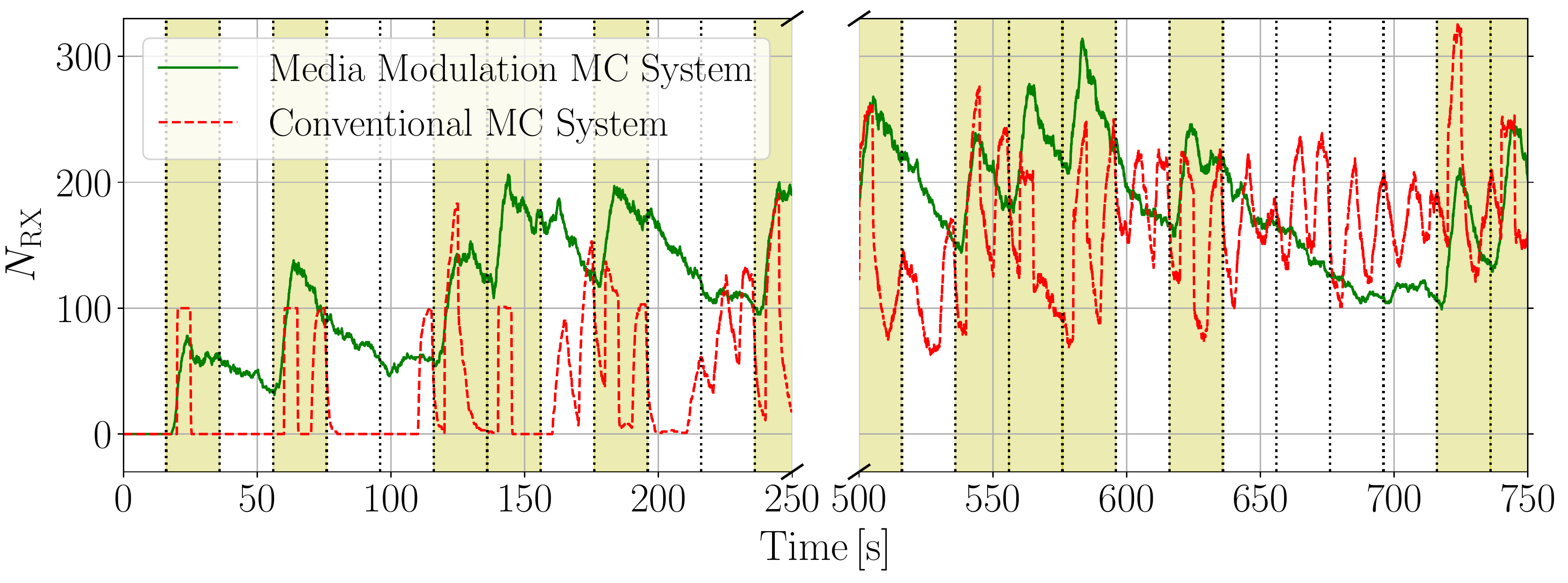}
    \caption{Received signal of the system using media modulation (green) and a conventional \ac{MC} system (red), respectively. For illustration, time intervals where binary symbol $1$ is transmitted are highlighted by yellow segments. Vertical dotted lines (black) indicate the symbol interval limits.
}\label{subfig-5}\vspace{-0.3cm}
  \end{figure}

\Figure{subfig-5} shows the number of observed blue \ac{SMs} at the \ac{RX} over time.
Yellow segments and vertical dotted black lines indicate transmissions of bit $1$ and the symbol interval boundaries, respectively.
For the media modulation system, we observe in \Figure{subfig-5} that bit $1$ transmissions are associated with sharp peaks of the number of \ac{SMs} counted at the \ac{RX} \textit{throughout the transmission}.
This is non-intuitive at first glance, as already from the beginning many \ac{SMs} are in the channel, i.e., the transmission medium appears crowded as can be observed in \Figure{fig:PBS_MM}.
However, as the \ac{EX} switches the \ac{SMs} from past transmissions to the non-observable state B, only the \ac{SMs} modulated for the current bit transmission are relevant.
This can be observed in the snapshot in the right side of \Figure{fig:PBS_MM}, where the \ac{SMs} of both states A and B enter the \ac{EX}, but leave the \ac{EX} only in state B, and hereby \ac{ISI} is avoided.
Furthermore, the \ac{TX} modulates \ac{SMs} only within its confined volume.
Hence, for each sent binary symbol $1$, a clear peak appears in the received signal.
In contrast, for the conventional \ac{MC} system, \Figure{subfig-5} shows clear peaks only in the beginning of the transmission.
For longer bit sequences, soiling occurs as \ac{SMs} from previous transmissions remain as waste in the channel.
In consequence, the molecules representing the current bit overlap with molecules from previous transmissions, which is visible also in the snapshot in the right side of \Figure{fig:PBS_classic}.
This overlap severely degrades the communication performance of the conventional \ac{MC} system.
In fact, for the conventional \ac{MC} system, we observe several peaks due to \ac{ISI} for $ 650 \, \si{\second} < t < 700 \, \si{\second}$ although there is no bit $1$ transmission.
From this, we conclude that for $t \rightarrow \infty$, only the media modulation based system allows for reliable information transmission.
Hence, media modulation facilitates \textbf{mitigation of \ac{ISI}}, which is crucial for long-term operation of synthetic \ac{MC} systems.

Furthermore, the media modulation system is more \textbf{resource efficient}. Although initially, i.e., before the first information transmission can begin, a large number of \ac{SMs} has to be deployed in the medium, no additional \ac{SMs} have to be added during transmission, and, hence, a lower number of \ac{SMs} is required in closed-loop systems employing media modulation compared to the conventional approach. This in consequence avoids the need for undesired large \ac{SM} storage units.

\subsection{Promising SMs for the Practical Implementation of Media Modulation}\label{MediaModulation}

In the following, we propose two types of switchable \ac{SMs}, which seem well suited for the practical realization of synthetic media modulation based \ac{MC} systems. The states of these \ac{SMs} can be externally read via fluorescence, which is a well established read-out method in \ac{MC} and is already used, e.g., in the testbed introduced in \cite{bartunik2021increasing}. Furthermore, using fluorescence allows for external detection. This makes \ac{SMs} with fluorescent properties well suited for healthcare applications, as, for detection, penetration into the channel is not needed.

Note that synthetic \ac{MC} systems, which employ phosphopeptides \cite{soldner2020survey} and redox reactions \cite{terrell2021bioelectronic}, have been discussed in the literature and may be interpreted as media modulation based systems due to their utilization of switchable \ac{SMs}. However, synthetic end-to-end implementations of these systems do not exist, yet. This may be due to the non-selectivity of redox reactions, i.e., redox reactions can be triggered unintentionally by molecules in the environment, and the high complexity of \ac{TX} and \ac{RX} design for phosphopeptides. Therefore, in this paper, we focus on fluorescent \ac{SMs}, but alternative options for switchable \ac{SMs} are of interest for future research.

\subsubsection{FlipGFP as a Switchable SM for Anomaly Detection}\label{FlipGFP}
FlipGFP is a one-time switchable \ac{SM}, cf. \Section{blankStateSignalingMolecules}. In particular, FlipGFP is a GFP-based fluorogenic protease reporter, i.e., FlipGFP can be used to detect proteases, which are enzymes supporting the process of breaking down proteins. One group of proteases are caspases. Caspases play an important role in cell apoptosis, i.e., programmed cell death after cell damage. High concentrations of caspases in natural systems, e.g., in the human cardiovascular system, may indicate anomalies. For example, a high concentration of caspase-$3$ in a given segment of the cardiovascular system may result from severe apoptosis provoked by a tumor. Consequently, reliable detection methods for caspases are desired. In \cite{zhang2019designing}, the authors show that FlipGFP can be designed to be very selective, i.e., sensitive to only one desired protease. The target protease is determined by the gene sequence used for the manufacturing process of the FlipGFP. This allows the design of different FlipGFPs for the detection of different proteases. In particular, in \cite{zhang2019designing}, the detection of tobacco etch virus (TEV) protease and caspase-$3$ is shown.

Detection is possible as FlipGFP exists in two states.
In particular, there exists a default non-active state with low fluorescence and an active state with high fluorescence, in the following denoted as state B and state A, respectively. FlipGFP switches from state B to state A after being triggered by a protease. FlipGFPs are one-time switchable \ac{SMs}, i.e., switching from state A to state B is not possible. In an anomaly detection task, the existence and the severity, i.e., the concentration of proteases, can be estimated by reading out the state distribution of the FlipGFPs by measuring the fluorescence intensity at the \ac{RX}. To estimate the proteases' origin, i.e., the location of an anomaly, e.g., a tumor, the propagation of the FlipGFPs within a given channel, e.g., inside a blood vessel, and the switching probability of the FlipGFPs have to be taken into account. As the switching process, i.e., the flipping from state B to state A, consists of several subprocesses \cite{zhang2019designing}, in-depth mathematical investigations are required for reliable and robust anomaly detection and estimation.

\subsubsection{GFPD as a Reversibly On-off Switchable SM for MC Systems}\label{GFPD_intro}

GFPD is a photochromic molecule, which can be used as \ac{SM} in closed-loop media modulation based \ac{MC} systems. Photochromic molecules can be reversibly switched between two states, where the transition between the states is induced by a light stimulus. Hereby, the state of a photochromic molecule reflects the embedded information. GFPD molecules have the key property that their fluorescence can be switched on and off by light stimuli of \textit{mutually different} wavelengths \cite{brakemann2011reversibly}, i.e., optical sources emitting light at different wavelengths can be used as \ac{TX} unit to modulate information (B $\rightarrow$ A) and \ac{EX} unit to delete the information (A $\rightarrow$ B). Moreover, a fluorescence detector equipped with a fluorescence-stimulating light source and an optical sensor can be employed as \ac{RX}. Furthermore, GFPD has been shown to be durable, i.e., stable over several photo-switching cycles \cite{brakemann2011reversibly}, which allows the long-term usage of GFPD as \ac{SM}. However, GFPD in state B can spontaneously switch to state A without any external trigger, which presents a GFPD specific challenge for information transmission applications.

\subsection{GFPD Based Media Modulation: Theoretical and First Experimental Results}\label{GFPD_theory_and_practice}

In this section, we provide a proof of concept for media modulation in \ac{MC} utilizing GFPD as switchable \ac{SM}.  First, in \Section{subsec:TheoreticalResults}, we present theoretical performance results to illustrate the feasibility of media modulation based MC using GFPD \cite{Brand2022MediaModulation}. Then, we depict an experimental setup for a communication link in a closed-loop environment, and show corresponding initial results in \Section{subsec:practicalResults}. These results demonstrate that reliable communication based on media modulation is possible in practice.

\subsubsection{Theoretical Performance Results for Media Modulation}\label{subsec:TheoreticalResults}

\begin{figure}[!tbp]
  \centering
  \includegraphics[width = 1\columnwidth, trim={0 0 0 1.3cm},clip]{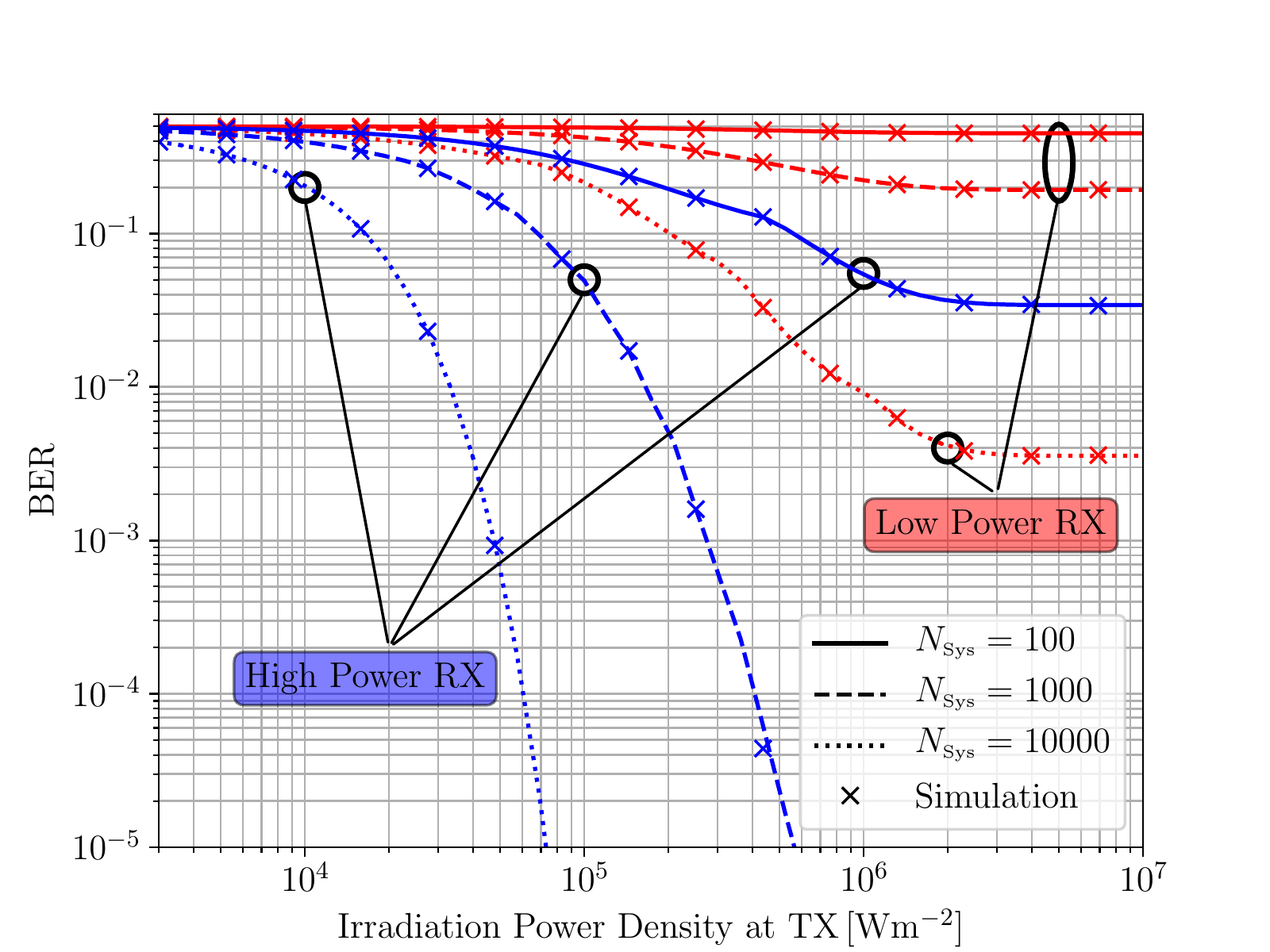}
  \caption{\acs{BER} as a function of the \ac{TX} irradiation power density for different numbers of signaling molecules $N_{\mathrm{Sys}}$ and two different \ac{RX} irradiation powers. In particular, 5 kW and 50 MW are used for the low power and the high power RX, respectively, which can for example be realized by lasers as shown in \cite{brakemann2011reversibly}. The \acs{BER} results are obtained for the system described in \cite{Brand2022MediaModulation}, which is similar to the one described in \Section{CompareBothSystems}, however, GFPD specific values are employed. Thus, the modulation process at the TX and the readout process at the RX become probabilistic, i.e., these processes are noisy. Furthermore, uniform flow was considered for analytical tractability; see \cite{Brand2022MediaModulation} for details on the default parameters and system model. The results obtained analytically and by Monte Carlo simulations are depicted by lines and markers, respectively.\vspace{-0.3cm}}
  \label{fig:BER}
\end{figure}

To enable media modulation, GFPD can be used as switchable \ac{SM} for information transmission in a three-dimensional duct system. The system comprises an \ac{EX}, a \ac{TX}, and an \ac{RX} for erasing, writing, and reading of information via external light, respectively, cf. upper right part of \Figure{fig:level_development}. The GFPD molecules are assumed to be already present in the channel. In contrast to conventional \ac{MC} systems, here, the statistical model for the received signal has to incorporate additional noise sources, which are media modulation and GFPD specific, e.g., the spontaneous state switching behavior, cf. \Section{GFPD_intro}. We discuss one media modulation specific noise source as an example in \Section{sec:theoretical_modeling}. Based on the statistical model, analytical expressions for the optimal threshold value of a threshold based detector and the resulting \ac{BER} can be derived \cite{Brand2022MediaModulation}.

\Figure{fig:BER} shows the \ac{BER} as a function of the irradiation power density at the TX for different numbers of \ac{SMs} and different RX radiation powers. We observe that the \ac{BER} decreases for increasing $N_{\mathrm{Sys}}$. This is intuitive, as for large $N_{\mathrm{Sys}}$, on average, more switchable \ac{SMs} are available at the TX site and switched during modulation. Therefore, more \ac{SMs} convey the information to be transmitted, which increases the reliability of the communication. Furthermore, we observe that increasing the powers of the light sources at TX and RX, respectively, decreases the \ac{BER}. This can be explained by the power dependent success probabilities of switching at the TX and the fluorescence based read-out at the RX, respectively, which are larger for higher powers. Note that for small \acp{BER} a minor deviation between results obtained analytically (dashed lines) and by Monte Carlo simulation (markers) is observed due to the slower convergence of simulation results to the theoretically expected ensemble average for low probability events. Finally, we observe that there exists a trade-off between $N_{\mathrm{Sys}}$ and the irradiation power, i.e., for smaller $N_{\mathrm{Sys}}$, the irradiation power has to be increased to achieve a certain \ac{BER}.

\subsubsection{Experimental Setup and First Successful Communication Employing GFPD as SM}\label{subsec:practicalResults}

\begin{figure}[!thp]
  \centering
  \includegraphics[width=1\columnwidth, trim={0cm 0 0cm 0},clip]{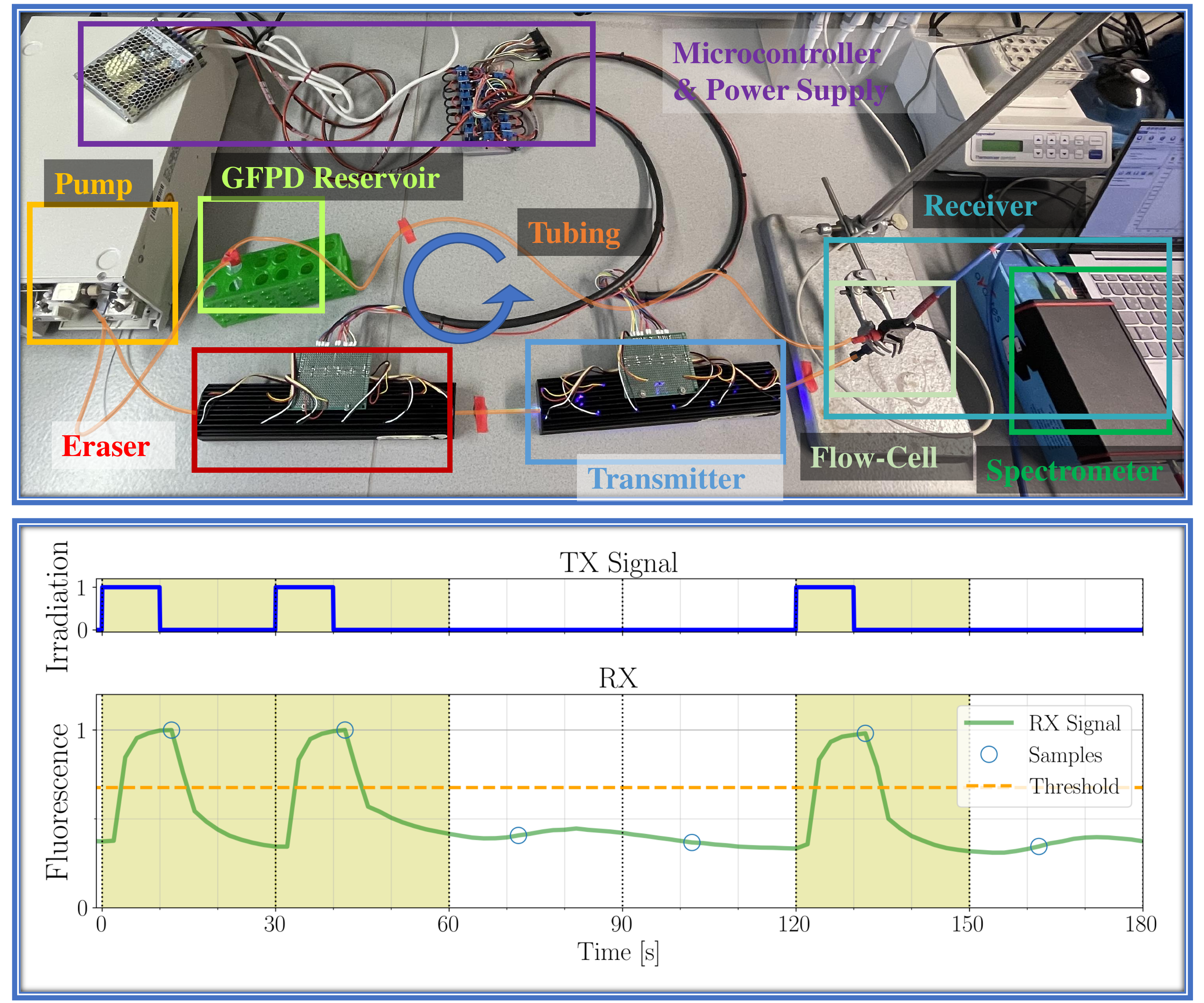}
  \caption{Top: Experimental closed-loop \ac{MC} setup for testing the end-to-end communication link using GFPD as \ac{SMs} and media modulation. The setup contains a reservoir, pump, connecting tubes, and illumination areas for \ac{EX}, \ac{TX}, and \ac{RX}. Bottom: Normalized transmit signal facilitating the media modulation process (solid blue) and fluorescence intensity at wavelength $\lambda_{\mathrm{F}} = 529 \, \si{\nm}$ (solid green) at the \ac{RX} showing clear peaks after bit 1 transmissions. The yellow shaded symbol intervals highlight bit 1 transmissions.\vspace{-0.3cm}}\label{fig:setup}
\end{figure}
We have conducted end-to-end \ac{MC} experiments, which validate the practical feasibility of media modulation.
The experimental setup comprises several building blocks, which are highlighted by colored boxes in \Figure{fig:setup}.
In the presented experiment, we use a total amount of $V = 6 \, \si{\milli \litre}$ of circulating fluid buffer, which contains $ 0.3 \, \si{\milli \gram \per \milli\litre}$ GFPD.
A flow pump generates an effective volume flux of $Q = 9.45 \, \si{\milli \litre \per \minute}$, which continuously moves the dissolved GFPD through a tube of total length $L_{\mathrm{T}} = 2.7 \, \si{\meter}$ and radius $0.6 \, \si{\milli \metre}$. The average loop duration $T_{\mathrm{L}}$ can be calculated to $T_{\mathrm{L}} = \frac{V}{Q} = 38.1 \, \si{\second}$.
The \ac{EX} and the \ac{TX} are implemented as irradiation modules of length $L_{\mathrm{I}} = 29 \, \si{\centi \metre}$, which comprise $12$ equally spaced light-emitting diodes (LEDs) able to emit light of wavelengths $\lambda_{\mathrm{OFF}} = 405 \, \si{\nm}$ and $\lambda_{\mathrm{ON}} = 365 \, \si{\nm}$, respectively.
At the \ac{RX}, at a distance of $6 \, \si{\centi\meter}$ from the \ac{TX}, a flow-cell equipped with one LED with wavelength $\lambda_{\mathrm{T}} = 500 \, \si{\nm}$ triggers the fluorescence of the GFPD, which is then recorded by a spectrometer with sampling frequency $f_{\mathrm{s}} = 0.5\, \si{\per \second}$.
In particular, after being triggered by light of wavelength $\lambda_{\mathrm{T}} = 500 \, \si{\nm}$, the dissolved GFPD molecules re-emit light of wavelength $\lambda_{\mathrm{F}} = 529 \, \si{\nm}$. The intensity of the re-emission hereby depends on the current states of the GFPD, which are controlled by the \ac{TX} during the modulation process.

In the lower part of \Figure{fig:setup}, we show the normalized (rectangular) irradiation signal used at the \ac{TX} to transmit bit sequence $\langle 1, 1, 0, 0, 1, 0 \rangle$ as well as the corresponding normalized fluorescence intensity at $\lambda_{\mathrm{F}} = 529 \, \si{\nm}$ measured at the \ac{RX}.
The symbol duration is set to $T_{\mathrm{S}} = 30 \, \si{\second}$, which results in a data rate of $\frac{1}{30} \, \si{\bit \per \second}$.
For bit 1 transmissions, the \ac{TX} irradiates the tube section for $10 \,\si{\second}$, followed by a guard interval (pause) of duration $T_{\mathrm{G}} = 20\, \si{\second}$, while for bit 0 transmissions, the \ac{TX} remains inactive.
The \ac{EX} is permanently active, and has been activated prior to the first transmission starting at $t=0$.
We observe that, as expected, a bit 1 transmission results in a clear fluorescence peak at the \ac{RX}, while during a bit 0 transmission the received fluorescence is nearly constant. Note that a minor increase can be seen for bit 0 transmissions at $t=80 \, \si{\second}$ and $t=170 \, \si{\second}$. This increase appears roughly $38 \,\si{\second}$ after the peak of the previous bit 1 transmission, which equals the loop duration $T_{\mathrm{L}}$. We conclude that this increase is caused by residual \ac{ISI}. Nevertheless, for the considered transmit bit sequence, error free detection by a simple static threshold-based detector is possible. Here, the first bit 1 transmission may act as a pilot signal to estimate the threshold value and the sampling times. In particular, the threshold value is calculated as the mean between the initial fluorescence and the value of the first peak. Moreover, the sampling times, which correspond to the expected peak times, are also derived from the pilot signal. From \Figure{fig:setup} we observe that the sample values (markers) are clearly located below (bit 0) or above (bit 1) the threshold (yellow dashed line), guaranteeing error-free transmission.
We further observe that the detected fluorescence before the transmission, i.e., at $t=0$, is larger than zero. From this observation, we conclude that the power of the \ac{EX} is not large enough to switch all GFPD molecules to the non-fluorescent state; this issue can be mitigated by utilizing brighter LEDs in future experiments. Nevertheless, the \ac{EX} power is already high enough to successfully suppress \ac{ISI} and enable reliable information transmission.

\scaleSubsection
\section{Directions for Future Research}\label{future_work}
\label{sec:Challenges}
\scaleSubsectionBelow
In this section, we briefly discuss some challenges arising in media modulation based \ac{MC}, which provide interesting future research opportunities.
\scaleSubsection
\subsection{Design, Manufacturing, and Testbed Implementation of Switchable SMs}\label{sec:practical_challenges}
\scaleSubsectionBelow
For media modulation, switchable \ac{SMs} are needed. We have presented two promising candidate \ac{SMs}, cf. \Section{MediaModulation}. However, further switchable \ac{SMs} are needed to suit different environments, e.g., for air-based \ac{MC}, and different constraints, e.g., bio-compatibility and possibly bio-degradability, which are crucial for \textit{in vivo} \ac{MC} systems. Furthermore, the development of meta molecules with multiple features, e.g., functionalized surface, magnetic core, self-destruction upon failure, etc., may lead synthetic \ac{MC} towards first \textit{in vivo} applications, and is therefore highly desirable.
Of course, the manufacturing process of these complex \ac{SMs} is challenging, which highlights the need for interdisciplinary cooperations with specialists from synthetic biology and bio-nanotechnology. In addition, as switchable \ac{SMs} pave the way to long-term experiments in testbeds, corresponding lab and communication protocols have to be established to ensure the reproducibility and comparability of results obtained with different testbeds, respectively.
\scaleSubsection
\subsection{Additional Noise Sources}\label{sec:theoretical_modeling}
\scaleSubsectionBelow
Due to the new concept of re-using available switchable \ac{SMs}, there are additional sources of noise, which impair the performance besides molecule propagation induced diffusion noise. These noise sources have to be identified and analyzed in order to optimize the RX design.

For example, for media modulation, the number of \ac{SMs} available at the modulation site at a given time can not be precisely controlled, i.e., it is not deterministic. Hence, to enable reliable transmission, knowledge of the probability distribution of the number of \ac{SMs} at the modulation site is needed, which may be acquired, e.g., via pilot signals.
\scaleSubsection
\subsection{Data-Driven Detection}\label{sec:data_driven_detection}
\scaleSubsectionBelow

Another challenge arises for more complex systems, e.g., media modulation based MC systems with cross-reactive \ac{SMs}, non-linear \ac{ISI}, or rare and unpredictable impairments during long-term transmission. Incorporating all possible sources of randomness into a comprehensive mathematical model may become challenging and eventually infeasible. Hence, for such systems, the application of data-driven methods for signal detection at the \ac{RX} may be beneficial \cite{huang2021signal}.

\section{Conclusion}
\label{sec:conclusion}
\scaleSectionBelow
In this article, switchable \ac{SMs} and their advantageous properties, which allow for writing information into their state by media modulation, have been discussed. In particular, different switchable \ac{SMs} have been investigated, such as one-time switchable and reversibly switchable \ac{SMs}, which overcome certain issues \ac{MC} currently faces and enable a number of future \ac{MC} applications. In addition, media modulation based \ac{MC} using GFPD as switchable \ac{SM} has been advocated and successfully validated with an experimental testbed. Finally, we have highlighted directions for future research with respect to the design, modeling, and manufacturing of switchable \ac{SMs}.

\bibliographystyle{IEEEtran}
\bibliography{literature}
\scaleSubsection
\section*{Acknowledgment}
\scaleSectionBelow
{
We thank Prof. Stefan Jakobs (Max Planck Institute for Biophysical Chemistry, Göttingen, Germany) for providing a plasmid encoding Dreiklang.
}
\scaleSubsection
\section*{Authors}
\scaleSectionBelow
{
\begin{description}
    \item[Lukas Brand] [S] is a Ph.D. student at the Institute for Digital Communications (IDC) at Friedrich-Alexander University (FAU) Erlangen-Nürnberg, Germany, with research focus on MC.
    \item[Maike Scherer] is a Ph.D. student at the Institute of Bioprocess Engineering at FAU with research focus on experimental implementation of MC systems.
    \item[Sebastian Lotter] [M] is a postdoctoral researcher at IDC at FAU with research focus on MC.
    \item[Teena tom Dieck] is a master student at FAU doing her research internship on MC at IDC.
    \item[Maximilian Schäfer] [M] is a postdoctoral researcher at IDC at FAU with research focus on MC.
    \item[Andreas Burkovski] received the Ph.D. degree in biology from the University of Osnabrück in 1993.
    Since 2002, he has been a Professor of microbiology at FAU.
    \item[Heinrich Sticht] received the Ph.D. degree in biochemistry from the University of Bayreuth in 1995.
    Since 2002, he has been a Professor of bioinformatics at FAU.
    \item[Kathrin Castiglione] received the Ph.D. degree in bioprocess engineering from the Technical University of Munich in 2009.
    Since 2018, she has been a Professor of bioprocess engineering at FAU.
    \item[Robert Schober] [F] received the Ph.D. degree in electrical engineering from FAU in 2000.
    Since 2012, he has been an Alexander von Humboldt Professor and head of IDC at FAU.
\end{description}
}
\end{document}